\documentclass{article}%
\usepackage{amssymb}
\usepackage{amsmath}
\usepackage{amsfonts}
\usepackage{graphicx}%
\newtheorem{theorem}{Theorem}

\newtheorem{corollary}[theorem]{Corollary}

\newtheorem{definition}[theorem]{Definition}

\newtheorem{proposition}[theorem]{Proposition}
\newtheorem{remark}[theorem]{Remark}

\newtheorem{assumption}[theorem]{Assumption}

\begin{document}

\title{Hylomorphic solitons for the generalized KdV equation.}
\author{Vieri Benci$^{\ast}$, Donato Fortunato$^{\ast\ast}$\\$^{\ast}$Dipartimento di Matematica\\Universit\`{a} degli Studi di Pisa \\Via Filippo Buonarroti 1/c, 56127 Pisa, Italy\\e-mail: benci@dma.unipi.it\\$^{\ast\ast}$Dipartimento di Matematica \\Universit\`{a} degli Studi di Bari Aldo Moro and\\INFN sezione di Bari\\Via Orabona 4, 70125 Bari, Italy\\e-mail: fortunat@dm.uniba.it}
\maketitle

\begin{abstract}
In this paper we prove the existence of hylomorphic solitons in the
generalized KdV equation. Following \cite{milan}, a soliton is called
hylomorphic if it is a solitary wave whose stability is due to a particular
relation between energy and another integral of motion which we call hylenic charge.

\bigskip

\textit{Dedicated to our friend Djairo De Figuereido on the occasion of his
80-th birthday}

\bigskip

AMS subject classification: 74J35, 35C08, 35A15, 35Q74, 35B35

\bigskip

Key words: Generalized KdV equation, solitary waves, hylomorphic solitons.

\end{abstract}
\tableofcontents

\section{ Introduction}

The Korteweg--de Vries equation (KdV) was first introduced by Boussinesq
(1877) and rediscovered by Diederik Korteweg and Gustav de Vries (1895). It is
a model of waves on shallow water surfaces.

Many different variations of the KdV equation have been studied. The most
common is the following one which is known as the generalized KdV equation
(gKdV):%
\begin{equation}
\frac{\partial u}{\partial t}+\frac{\partial^{3}u}{\partial x^{3}}%
-\frac{\partial}{\partial x}W^{\prime}(u)=0 \label{1K}%
\end{equation}
where $u=u(t,x),\ $and $W\in C^{2}(\mathbb{R)}$. If $W(s)=-s^{3},$ then
(\ref{1K}) reduces to the usual KdV equation
\begin{equation}
\frac{\partial u}{\partial t}+\frac{\partial^{3}u}{\partial x^{3}}%
+6u\frac{\partial u}{\partial x}=0 \label{1+K}%
\end{equation}
If $W(s)=-\frac{s^{k+2}}{\left(  k+2\right)  \left(  k+1\right)  },$ then
(\ref{1K}) reduces to the equation%
\begin{equation}
\frac{\partial u}{\partial t}+\frac{\partial^{3}u}{\partial x^{3}}+u^{k}%
\frac{\partial u}{\partial x}=0 \label{mkdv}%
\end{equation}
known as the modified KdV equation (mKdV).

In this paper we are interested to the existence of solitary waves and
solitons for the gKdV equation. Roughly speaking a \textit{solitary wave} is a
solution of a field equation whose energy travels as a localized packet and
which preserves this localization in time. A \textit{soliton} is a solitary
wave which exhibits some form of stability so that it has a particle-like behavior.

Using the inverse scattering transform, it is possible to prove that KdV
admits soliton solutions and to have an extremely powerful and precise
information on them. However the inverse scattering techniques cannot be
applied to the generalized KdV equation. In this paper, we shall use the
method developed in \cite{befolib} to prove that equation (\ref{1K}) admits
solitons and we will show that they are \textit{hylomorphic}. Following
\cite{milan}, a soliton is called \textit{hylomorphic} if its stability is due
to a particular interplay between the \textit{energy }$E$ and the
\textit{hylenic} \textit{charge }$C:=\int u^{2}dx$ which is another integral
of motion. More precisely, a soliton $u_{0}$ is hylomorphic if%
\[
E(u_{0})=\min\left\{  E(u)\ |\ \int u^{2}dx=C(u_{0})\right\}  .
\]
We will show that eq. (\ref{1K}) admits solitons provided that $W$ satisfies
suitable assumptions (see Theorem \ref{imp2K}). In particular, if
$W(s)=-\frac{s^{k+2}}{\left(  k+2\right)  \left(  k+1\right)  }$, hylomorphic
solitons exist for $k=1,2,3.$ So, in this case, we get a different proof of
well known results (see \cite{We86}) and we show that the "usual" solitons of
mKdV can be considered "hylomorphic".

In Th. \ref{imp2K}, we obtain the existence of hylomorphic solitons under a
very general set of assumptions on $W;$ moreover, in contrast to other results
on this topic, these assumptions are easy to verify.

\section{Solitary waves and solitons\label{sws}}

In this section we construct a functional abstract framework which allows to
define solitary waves, solitons and hylomorphic solitons.

\subsection{Solitary waves\label{be}}

\textit{Solitary waves} and solitons are particular \textit{states} of a
dynamical system described by one or more partial differential equations.
Thus, we assume that the states of this system are described by one or more
\textit{fields} which mathematically are represented by functions
\[
\mathbf{u}:\mathbb{R}^{N}\rightarrow V
\]
where $V$ is a vector space with norm $\left\vert \ \cdot\ \right\vert _{V}$
and which is called the internal parameters space. We assume the system to be
deterministic; this means that it can be described as a dynamical system
$\left(  X,\gamma\right)  $ where $X$ is the set of the states and
$\gamma:\mathbb{R}\times X\rightarrow X$ is the time evolution map. If
$\mathbf{u}_{0}(x)\in X,$ the evolution of the system will be described by the
function
\begin{equation}
\mathbf{u}\left(  t,x\right)  :=\gamma_{t}\mathbf{u}_{0}(x). \label{flusso}%
\end{equation}
We assume that the states of $X$ have "finite energy" so that they decay at
$\infty$ sufficiently fast and that%
\begin{equation}
X\subset L_{loc}^{1}\left(  \mathbb{R}^{N},V\right)  . \label{lilla}%
\end{equation}

Thus we are lead to give the following definition:\label{pag}

\begin{definition}
\label{ft}A dynamical system $\left(  X,\gamma\right)  $ is called of FT type
(field-theory-type) if $X$ is a Hilbert space of functions of type
(\ref{lilla}).
\end{definition}

Let $\mathcal{T}$ be the group of translations in $\mathbb{R}^{N}$ and
$U\left(  V\right)  $ the group of unitary transformations on $V;$ set
\[
G=\mathcal{T}\times U\left(  V\right)
\]

Given $\left(  \tau,h\right)  \in G$ we will consider the representation of
$G$ on $X\subset L_{loc}^{1}\left(  \mathbb{R}^{N},V\right)  $ given by%
\[
\left[  T_{\left(  \tau,h\right)  }\mathbf{u}\right]  (x)=h\mathbf{u}(x-\tau)
\]
For example take $X=L^{2}\left(  \mathbb{R}^{N},\mathbb{C}\right)
,\ h=e^{i\theta}\in U(1),$ then%
\[
\left[  T_{\left(  \tau,h\right)  }u\right]  (x)=e^{i\theta}u(x-\tau)
\]

A solitary wave is a state of finite energy which evolves without changing his
shape. This informal description can be formalized by the following definition:

\begin{definition}
\label{solw} A state $\mathbf{u}_{0}\in X\backslash\left\{  0\right\}  ,$ is
called solitary wave if there is a continuous trajectory
\[
t\mapsto\left(  \tau(t),h(t)\right)  \in G
\]
such that%
\[
\gamma_{t}\mathbf{u}_{0}(x)=h(t)\mathbf{u}_{0}(x-\tau(t))
\]

\end{definition}

\bigskip

For example, consider a solution of a field equation having the following
form:
\begin{equation}
\mathbf{u}\left(  t,x\right)  =u_{0}(x-vt-x_{0})e^{i(v\mathbf{\cdot
}x\mathbf{-}\omega t)};\ u_{0}\in L^{2}(\mathbb{R}^{N}); \label{solwav}%
\end{equation}
$x_{0},v\in\mathbb{R}^{N},\omega\in\mathbb{R}.\ $Clearly $\mathbf{u}\left(
t,x\right)  $ is a solitary wave for every $t\in\mathbb{R}$. The evolution of
a solitary wave is a translation plus a unitary change of the internal
parameters (in this case the phase).

\bigskip

\subsection{Orbitally stable states and solitons}

The \textit{solitons%
\index{soliton}%
} are solitary waves characterized by some form of stability. To define them
at this level of abstractness, we need to recall some well known notions in
the theory of dynamical systems.

\begin{definition}
A set $\Gamma\subset X$ is called \textit{invariant} if $\forall\mathbf{u}%
\in\Gamma,\forall t\in\mathbb{R},\ \gamma_{t}\mathbf{u}\in\Gamma.$
\end{definition}

\begin{definition}
Let $\left(  X,\gamma\right)  $ be a dynamical system and let $X$ be equipped
with a metric $d$ (it is not necessary to assume that $d(\mathbf{u,v}%
)=\left\Vert \mathbf{u}-\mathbf{v}\right\Vert _{X}$). An invariant set
$\Gamma\subset X$ is called stable (with respect to $d$), if $\forall
\varepsilon>0,$ $\exists\delta>0,\;\forall\mathbf{u}\in X$,
\[
d(\mathbf{u},\Gamma)\leq\delta,
\]
implies that
\[
\forall t\geq0,\text{ }d(\gamma_{t}\mathbf{u,}\Gamma)\leq\varepsilon.
\]

\end{definition}

\begin{definition}
\label{carla}Let $\left(  X,d\right)  $ be a metric space and let
$\mathcal{T}$ be the group of translations. A set $\Gamma\subset X$ is called
$\mathcal{T}$-compact if for any sequence $\mathbf{u}_{n}(x)\in\Gamma\ $there
is a subsequence $\mathbf{u}_{n_{k}}$ and a sequence $\tau_{k}\in\mathcal{T}$
such that $\mathbf{u}_{n_{k}}(x-\tau_{k})$ is convergent with respect to the
metric $d$.
\end{definition}

Now we give the definition of \textit{orbitally stable state}:

\begin{definition}
\label{dos}Let $\left(  X,\gamma\right)  $ be a dynamical system with $X$
equipped with a metric $d$. A state $\mathbf{u}\in X\mathbf{\ }$ is called
\textit{orbitally stable} (with respect to $d$) if $\mathbf{u}\in\Gamma\subset
X$ where

\begin{itemize}
\item (i) $\Gamma$ is an invariant stable set with respect to $d$,

\item (ii) $\Gamma$ is $\mathcal{T}$-compact (with respect to $d$).
\end{itemize}
\end{definition}

This definition is usually present in the literature relative to the dynamics
of PDE's (see e.g. \cite{Cazli}, \cite{We86} etc.).

Now we are able to give the definition of soliton:

\begin{definition}
\label{ds}Let $\left(  X,\gamma\right)  $ be a dynamical system with $X$
equipped with a metric $d.$ A soliton is an orbitally stable solitary wave
(with respect to $d$).
\end{definition}

\bigskip

\begin{remark}
In our definition, since $\left(  X,\gamma\right)  $ is a dynamical system,
the map%
\[
t\mapsto\gamma_{t}\mathbf{u}%
\]
is continuos with respect to $\left\Vert \mathbf{\cdot}\right\Vert _{X}$. In
the above definitions, we have introduced a distance $d,$ however we have not
supposed that
\[
d(\mathbf{u,v})=\left\Vert \mathbf{u}-\mathbf{v}\right\Vert _{X}%
\]
In fact, in some applications this is not true. As we will see in section
\ref{HSK}, this is the case for equation (\ref{1K}) where we have%
\[
d(\mathbf{u,v})\leq M\left\Vert \mathbf{u}-\mathbf{v}\right\Vert _{X}%
\]
for a suitable constant $M>0.$
\end{remark}

\bigskip

\subsection{Hylomorphic solitons}

We now assume that the dynamical system $\left(  X,\gamma\right)  $ has two
constants of motion: the energy $E$ and the hylenic charge $C.$ At this level
of abstraction, of course, the name energy and hylenic charge are conventional.

\begin{definition}
\label{tdc}Let $\left(  X,\gamma\right)  $ be a dynamical system where $X$ is
equipped with a metric $d.$ A soliton $\mathbf{u}_{0}\in X$ is called
\textbf{hylomorphic }if the set $\Gamma$ (given by Def. \ref{dos}) has the
following structure%
\begin{equation}
\Gamma=\Gamma\left(  e_{0},c_{0}\right)  =\left\{  \mathbf{u}\in
X\ |\ E(\mathbf{u})=e_{0},\ \left\vert C(\mathbf{u})\right\vert =c_{0}%
\right\}  \label{plis}%
\end{equation}
where%
\begin{equation}
e_{0}=\min\left\{  E(\mathbf{u})\ |\ \left\vert C(\mathbf{u})\right\vert
=c_{0}\right\}  . \label{minbis}%
\end{equation}

\end{definition}

Notice that, by (\ref{minbis}), we have that a hylomorphic soliton
$\mathbf{u}_{0}$ minimizes the energy on%
\begin{equation}
\mathfrak{M}_{c_{0}}=\left\{  \mathbf{u}\in X\ |\ \left\vert C(\mathbf{u}%
)\right\vert =c_{0}\right\}  .
\end{equation}
If $\mathfrak{M}_{c_{0}}$ is a manifold and $E$ and $C$ are differentiable,
then $\mathbf{u}_{0}$ satisfies the following nonlinear eigenvalue problem:%
\[
E^{\prime}(\mathbf{u}_{0})=\lambda C^{\prime}(\mathbf{u}_{0}).
\]

\section{Hylomorphic solitons for the nonlinear Schr\"{o}dinger
equation\label{NSE}}

The solitons for eq. (\ref{1K}), as we will see, are related to the solitons
of the nonlinear Schr\"{o}dinger equation. We recall that the orbital
stability for the nonlinear Schr\"{o}dinger equation has been proved in
\cite{Cazli} (see also \cite{BBGM} for the general case and \cite{sulesulem}
with its references).

Here we shall use a method to prove the existence of hylomorphic solitons for
(\ref{1K}) similar to the one presented in \cite{befolib} (see also
\cite{befolak} and \cite{befobrez}). In this section we will resume this method.

The nonlinear Schr\"{o}dinger equation is given by
\begin{equation}
i\frac{\partial\psi}{\partial t}=-\frac{1}{2}\Delta\psi+\frac{1}{2}W^{\prime
}(\psi)\label{NSV}%
\end{equation}
where $\psi:\mathbb{R\times R}\rightarrow\mathbb{C\ }$and where
$W:\mathbb{C\rightarrow R}$ and
\begin{equation}
W^{\prime}(\psi)=\frac{\partial W}{\partial\psi_{1}}+i\frac{\partial
W}{\partial\psi_{2}}.\label{w'}%
\end{equation}
We assume that $W$ depends only on $\left\vert \psi\right\vert $, namely
\[
W(\psi)=F(\left\vert \psi\right\vert )\ \text{and so\ }W^{\prime}%
(\psi)=F^{\prime}(\left\vert \psi\right\vert )\frac{\psi}{\left\vert
\psi\right\vert }.
\]
for some smooth function $F:\left[  0,\infty\right)  \rightarrow\mathbb{R}.$
In the following we shall identify, with some abuse of notation, $W$ with $F.$

The energy is given by.
\begin{equation}
E=\int\left(  \frac{1}{2}\left\vert \nabla\psi\right\vert ^{2}+W(\psi)\right)
dx \label{ghj}%
\end{equation}
Moreover the Schr\"{o}dinger equation has an other important integral of
motion%
\begin{equation}
C=\int\left\vert \psi\right\vert ^{2}dx \label{chery}%
\end{equation}
to which we will refer as\textit{ charge.}

We make the following assumptions on the function $W:$%
\begin{equation}
W(0)=W^{\prime}(0)=0 \label{Wa}%
\end{equation}%
\begin{equation}
W^{\prime\prime}(0)=2E_{0}>0 \label{Wb}%
\end{equation}
if we set
\begin{equation}
W(s)=E_{0}s^{2}+N(s), \label{W}%
\end{equation}
then,
\begin{equation}
\exists s_{0}\in\mathbb{R}^{+}\text{ such that }N(s_{0})<0 \label{W1}%
\end{equation}
there exist $q,$ $r$ in $(2,+\infty),$ s. t.%
\begin{equation}
|N^{\prime}(s)|\leq c_{1}s^{r-1}+c_{2}s^{q-1} \label{Wp}%
\end{equation}%
\begin{equation}
N(s)\geq-cs^{p},\text{ }c\geq0,\ 2<p<6\text{ for }s\text{ large} \label{W0}%
\end{equation}

We can apply the abstract theory of section \ref{sws} setting:

\begin{itemize}
\item $X=H^{1}(\mathbb{R},\mathbb{C}),$ $\mathbf{u}=\psi;$

\item $d(\psi,\varphi)=\left\Vert \psi-\varphi\right\Vert _{H^{1}}.$
\end{itemize}

\begin{theorem}
\label{solitoni}Let $W$ satisfy (\ref{Wa}),...,(\ref{W0}). Then there exists
$\delta_{\infty}>0$ such that for every $\delta\in\left(  0,\delta_{\infty
}\right)  $ there exist $c_{\delta}>0$ and an orbitally stable state
$\psi_{\delta}\in H^{1}(\mathbb{R},\mathbb{C}),$ such that $\psi_{\delta}$
minimizes the energy on the manifold%
\[
\mathfrak{M}_{c_{\delta}}=\left\{  \mathbf{u}\in X\ |\ \int\left\vert
\psi\right\vert ^{2}dx=c_{\delta}\right\}  .
\]
Moreover if $\delta_{1}<\delta_{2}$ we have that $c_{\delta_{1}}>c_{\delta
_{2}}$.
\end{theorem}

\textbf{Proof: }The proof is an immediate consequence of Th. 52 in
\cite{befolib} (see also \cite{befolak}).

$\square$

By the above theorem we have that every $\psi_{\delta}$ is an orbitally stable
state; the following theorem shows that it is a soliton.\bigskip

\begin{theorem}
\label{marina}Let $u_{\delta}\in H^{1}(\mathbb{R},\mathbb{C})$ be a orbitally
stable state as in Th. \ref{solitoni}. Then $u_{\delta}$ is a solution of the
equation%
\begin{equation}
-\frac{1}{2}\Delta u+\frac{1}{2}W^{\prime}(u)=\omega u\label{brb}%
\end{equation}
and
\begin{equation}
\psi_{\delta}\left(  t,x\right)  :=u_{\delta}(x)e^{-i\omega t}\label{lula2}%
\end{equation}
solves (\ref{NSV}). Namely $u_{\delta}$ is a (hylomorphic) soliton.
\end{theorem}

\textbf{Proof}. See Proposition 59 of \cite{befolib} (see also \cite{befolak}).

$\square$

\section{Hylomorphic solitons for the generalized KdV equation\label{HSK}}

\bigskip

First of all let us show that the "good" solutions of equation (\ref{1K}) have
two constants of motion.

\begin{proposition}
Let $W$ be a $C^{2}$ function and $u$ be a smooth solution of equation
(\ref{1K}) and assume that $u(t,.)\in H^{1}(\mathbb{R}),\frac{\partial
u}{\partial t}(t,\cdot)\in L^{2}(\mathbb{R}).$Then $u$ has two integrals of
motion: the energy
\begin{equation}
E=\int\left(  \frac{1}{2}\left[  \frac{\partial u}{\partial x}\right]
^{2}+W(u)\right)  dx \label{ebis}%
\end{equation}
and the charge%
\begin{equation}
C=\frac{1}{2}\int u^{2}dx \label{cbis}%
\end{equation}

\end{proposition}

\textbf{Proof.} Since $\frac{\partial u}{\partial t}(t,\cdot)\in
L^{2}(\mathbb{R})$ and
\[
-\frac{\partial^{3}u}{\partial x^{3}}+\frac{\partial}{\partial x}W^{\prime
}(u)=\frac{\partial u}{\partial t},
\]
the integral
\[
\int\left(  -\frac{\partial^{2}u}{\partial x^{2}}+W^{\prime}(u)\right)
\frac{\partial u}{\partial t}dx
\]
is well defined and it equals the time derivative of $E(u(t)):$
\begin{equation}
\frac{d}{dt}E(u(t))=\int\left(  -\frac{\partial^{2}u}{\partial x^{2}%
}+W^{\prime}(u)\right)  \frac{\partial u}{\partial t}dx \label{sa}%
\end{equation}

Moreover
\begin{equation}
\frac{\partial u}{\partial t}=-\frac{\partial^{3}u}{\partial x^{3}}%
+\frac{\partial}{\partial x}(W^{\prime}(u))=-\frac{\partial}{\partial
x}\left(  \frac{\partial^{2}u}{\partial x^{2}}-W^{\prime}(u)\right)
\label{se}%
\end{equation}
Substituting (\ref{se}) in (\ref{sa}), we get%
\begin{align*}
\frac{d}{dt}E(u)  &  =\int\left(  \frac{\partial^{2}u}{\partial x^{2}%
}-W^{\prime}(u)\right)  \frac{\partial}{\partial x}\left(  \frac{\partial
^{2}u}{\partial x^{2}}-W^{\prime}(u)\right)  dx\\
&  =\frac{1}{2}\int\frac{\partial}{\partial x}\left[  \left(  \frac
{\partial^{2}u}{\partial x^{2}}-W^{\prime}(u)\right)  ^{2}\right]  dx=0
\end{align*}
Then $E$ is constant along the solution $u.$

Let us now show that also $C$ is constant along $u.$ By (\ref{se}) we have%
\begin{equation}
\frac{d}{dt}C(u)=\int u\frac{\partial u}{\partial t}dx=\int u\left(
-\frac{\partial^{3}u}{\partial x^{3}}+\frac{\partial}{\partial x}W^{\prime
}(u)\right)  dx \label{si}%
\end{equation}
Let us compute each piece separately:%
\begin{equation}
\int u\left(  -\frac{\partial^{3}u}{\partial x^{3}}\right)  dx=\int
\frac{\partial u}{\partial x}\left(  \frac{\partial^{2}u}{\partial x^{2}%
}\right)  dx=\frac{1}{2}\int\frac{\partial}{\partial x}\left(  \frac{\partial
u}{\partial x}\right)  ^{2}=0 \label{su}%
\end{equation}
Moreover%
\begin{equation}
\int u\frac{\partial}{\partial x}(W^{\prime}(u))dx=-\int W^{\prime}%
(u)\frac{\partial u}{\partial x}dx=-\int\frac{\partial}{\partial x}W(u)dx=0
\label{so}%
\end{equation}

Substituting (\ref{so}) and (\ref{su}) in (\ref{si}) we get%
\[
\frac{d}{dt}C(u)=0
\]
$\square$

\bigskip

We will apply the abstract theory of section \ref{sws} setting:

\begin{itemize}
\item $X=H^{2}(\mathbb{R});$

\item $d(u,v)=\left\Vert \psi-\varphi\right\Vert _{H^{1}}.$
\end{itemize}

To this end, we need the following assumption which guarantees that also the
weak solutions in $H^{2}(\mathbb{R})$ have the properties required by the theory.

\begin{assumption}
\label{claim} \ We assume that the equation (\ref{1K}) defines a dynamical
system on $X=H^{2}(\mathbb{R})$, namely, for any initial data $u_{0}\in
H^{2}(\mathbb{R})$ there is a unique (weak) solution in $C(\mathbb{R}%
,H^{2}(\mathbb{R}))$ of the Cauchy problem. Moreover we assume that the energy
(\ref{ebis}) and the charge (\ref{cbis}) are conserved integrals.
\end{assumption}

\bigskip

\begin{remark}
\label{rem}Clearly, assumption \ref{claim} depends on $W.$ By the existence
theory of Kato \cite{kato}, the assumption%
\begin{equation}
\underset{s\rightarrow\pm\infty}{\lim\sup}\frac{-W^{\prime\prime}(s)}{s^{4}%
}\leq0 \label{base}%
\end{equation}
implies the existence of a unique global solution of eq. (\ref{1K}) in
$C(\mathbb{R},H^{2}(\mathbb{R}))$ and the conservation of (\ref{ebis}) and
(\ref{cbis}). In particular, if $W=-|u|^{k+2}$ we need $k<4$. (for the well
posedeness of equation (\ref{1K}) see also \cite{tsu70}, \cite{k93} and their references).
\end{remark}

\bigskip

\begin{theorem}
\label{theoKdV}Let $W$ satisfy the assumptions (\ref{Wa}),...,(\ref{W0}). Then
there exists $\delta_{\infty}>0$ such that for every $\delta\in\left(
0,\delta_{\infty}\right)  $ there exist $c_{\delta}>0$ and $u_{\delta}\in
H^{2}(\mathbb{R})$ which minimizes the energy $E$ on the manifold%
\[
\mathfrak{M}_{c_{\delta}}=\left\{  u\in X\ |\ \int u^{2}dx=c_{\delta}\right\}
.
\]
If $\delta_{1}<\delta_{2}$ we have that $c_{\delta_{1}}>c_{\delta_{2}}$.
Moreover, if also assumption \ref{claim} holds, then $u_{\delta}$ is an
orbitally stable state.
\end{theorem}

\textbf{Proof}: The proof of this theorem is essentially the same than the
proof of th. \ref{solitoni} which can be found in \cite{befolib}, Th. 52 (see
also \cite{befolak}). The reason for this relies on the fact that the energy
and the charge for eq. (\ref{NSV}) given by (\ref{ghj}) and (\ref{chery}) are
formally the same than the energy and the charge of eq. (\ref{1K}) given by
(\ref{ebis}) and (\ref{cbis}). The fact that in the first case $\psi$ is
complex while in the second case $u$ is real-valued does not affect the estimates.

Another difference concerns the space $X$ which is $H^{1}(\mathbb{R}%
,\mathbb{C})$ for eq. (\ref{NSV}) and $H^{2}(\mathbb{R})$ for eq. (\ref{1K}).

The proof of the theorem consists in minimizing a suitable functional
$K_{\delta}$ on $X$ (see (2.43) in \cite{befolib})$.$ In the case of eq.
(\ref{1K}), we minimize first the functional $K_{\delta}$ on $H^{1}%
(\mathbb{R})$ and then we can prove that the set $\Gamma_{\delta}$ of these
minimizers is contained in $H^{2}(\mathbb{R}).$ In fact the minimizers satisfy
the following eigenvalue equation
\[
-\frac{\partial^{2}u}{\partial x^{2}}+W^{\prime}(u)=\lambda_{\delta}u
\]
and hence, since $W\in C^{2},$ by the standard elliptic regularization, we
have that $\Gamma_{\delta}\subset H^{2}(\mathbb{R}).$

\bigskip$\square$

\begin{remark}
Assumption (\ref{Wb}) is not necessary. It is not restrictive to assume that%
\[
E_{0}>0
\]

\end{remark}

\textbf{Proof}. In fact consider the following equations
\begin{equation}
\frac{\partial u}{\partial t}+\frac{\partial^{3}u}{\partial x^{3}}%
-\frac{\partial}{\partial x}W_{0}^{\prime}(u)=0 \label{aaa}%
\end{equation}
where $W_{0}^{\prime\prime}(0)=-2E_{0}<0$. In this case is convenient to
consider the equation%
\begin{equation}
\frac{\partial v}{\partial t}+\frac{\partial^{3}v}{\partial x^{3}}%
-\frac{\partial}{\partial x}W^{\prime}(v)=0 \label{bbb}%
\end{equation}
where $W(s)=W_{0}(s)+\left(  E_{0}+1\right)  s^{2}.$We have that
\[
W^{\prime\prime}(0)=2>0
\]
and to every solution $v$ of eq. (\ref{bbb}) corresponds a solution%
\[
u(t,x)=v(t,x+ct)\ \ with\ \ c=2\left(  E_{0}+1\right)
\]
of eq. (\ref{aaa}). In fact%
\begin{align*}
\frac{\partial u}{\partial t}+\frac{\partial^{3}u}{\partial x^{3}}%
-\frac{\partial}{\partial x}W_{0}^{\prime}(u)  &  =\frac{\partial v}{\partial
t}+c\frac{\partial v}{\partial x}+\frac{\partial^{3}v}{\partial x^{3}}%
-\frac{\partial}{\partial x}W_{0}^{\prime}(v)\\
&  =\frac{\partial v}{\partial t}+c\frac{\partial v}{\partial x}%
+\frac{\partial^{3}v}{\partial x^{3}}-\frac{\partial}{\partial x}\left[
W^{\prime}(v)-2\left(  E_{0}+1\right)  v\right] \\
&  =\frac{\partial v}{\partial t}+\frac{\partial^{3}v}{\partial x^{3}}%
-\frac{\partial}{\partial x}W^{\prime}(v)+c\frac{\partial v}{\partial
x}-2\left(  E_{0}+1\right)  \frac{\partial v}{\partial x}\\
&  =0
\end{align*}

$\square$

We shall prove that the minimizer $u_{\delta}$ in Theorem \ref{theoKdV} is a
soliton (def. \ref{ds}). 

\begin{theorem}
\label{imp2K}Under the assumptions and the notations of Th. \ref{theoKdV}, the
minimizer $u_{\delta}$ is a (hylomorphic) soliton. Moreover it is a solution
of the equation%
\begin{equation}
\frac{\partial^{3}u_{\delta}}{\partial x^{3}}-\frac{\partial}{\partial
x}W^{\prime}(u_{\delta})=c_{\delta}\frac{\partial u_{\delta}}{\partial
x}\label{sat}%
\end{equation}
and%
\[
U_{\delta}(t,x):=u_{\delta}(x-c_{\delta}t)
\]
solves (\ref{1K}).
\end{theorem}

\textbf{Proof of Theorem \ref{imp2K}. }By th. \ref{theoKdV}, the minimizer
$u_{\delta}$ is an orbitally stable state. So, in order to show that it is a
soliton (def.\ref{ds}), we need to prove that $u_{\delta}$ is a solitary wave
(def.\ref{solw}).

Since $u_{\delta}$ is a minimizer of the energy $E$ on the manifold
$\mathfrak{M}_{c_{\delta}},$ there exists a Lagrange multiplier $c_{\delta}$ s.t.%

\[
E^{\prime}(u_{\delta})=-c_{\delta}C^{\prime}(u_{\delta}).
\]
The above equality can be written as follows%

\[
-\frac{\partial^{2}u_{\delta}}{\partial x^{2}}+W^{\prime}(u_{\delta
})=-c_{\delta}u_{\delta}%
\]
So, if we take the derivative $\frac{\partial}{\partial x}$ on both side, we
get (\ref{sat}). Finally (\ref{sat}) implies that the travelling wave
$u(t,x)=u_{\delta}(x-c_{\delta}t)$ solves (\ref{1K}) and consequently
$u_{\delta}$ is a solitary wave.

$\square$

\begin{corollary}
Equation (\ref{mkdv}) admits hylomorphic solitons for $k=1,2,3.$
\end{corollary}

\textbf{Proof:} Take%
\begin{equation}
W(s)=-\frac{s^{k+2}}{\left(  k+2\right)  \left(  k+1\right)  }. \label{finale}%
\end{equation}

For $k=1,2,3$ the function $W$ satisfies (\ref{Wa},...\ref{W0}) and
(\ref{base}). So, by Remark \ref{rem} also the assumption \ \ref{claim} is
satisfied. Then, by Theorem \ref{imp2K}, equation (\ref{mkdv}), for $k=1,2,3$,
admits hylomorphic solitons.

$\square$

\begin{corollary}
If $W(s)=-|s|^{k+2},$ then equation (\ref{1K}) admits hylomorphic solitons for
$k\in\left(  0,4\right)  .$
\end{corollary}

\textbf{Proof:} The proof is the same as for the above corollay.

$\square$


\begin{thebibliography}{99}                                                                                               %


\bibitem {BBGM}\textsc{J.Bellazzini, V.Benci, M.Ghimenti, A.M.Micheletti,}
\emph{\ On the existence of the fundamental eigenvalue of an elliptic problem
in $\mathbb{R}^{N\ }$ }, Adv. Nonlinear Stud. \textbf{7} (2007), 439--458

\bibitem {milan}\textsc{V.Benci, }\textit{Hylomorphic solitons, }Milan J.
Math., \textbf{77 }(2009), 271-332.

\bibitem {befogranas}\textsc{V. Benci, D. Fortunato,} \emph{Solitary waves in
the nonlinear wave equation and in gauge theories}, J. Fixed Point Theory
Appl. \textbf{1} (2007), 61-86

\bibitem {befolak}\textsc{V. Benci, D. Fortunato}, \textit{A minimization
method and applications to the study of solitons}, Nonlinear Anal. T. M.A.,
\textbf{75}, (2012), 4398-4421.

\bibitem {befolib}\textsc{V.Benci, D.Fortunato, }\textit{Variational methods
in nonlinear field equations, }Springer Monographs in Mathematics, Springer,
(2014), ISBN: 3319069136

\bibitem {befobrez}\textsc{V.Benci, D.Fortunato, }\textit{Solitons in
Schr\"{o}dinger-Maxwell equations, }J. Fixed Point Theory Appl. \textbf{15} (2014).

\bibitem {Cazli}\textsc{T.Cazenave, P.L.Lions, }\textit{Orbital stability of
standing waves for some nonlinear Schr\"{o}dinger equations, }Comm. Math.
Phys. \textbf{85 }(1982), 549-561.

\bibitem {evans}\textsc{C.Evans, }\textit{Partial differential equations}

\bibitem {kato}\textsc{T.Kato, }\textit{On the Cauchy problem for the
(generalized) Korteweg-de Vries equation}, Studies in applied mathematics,
Adv. Math. Suppl. Stud., vol. 8, Academic Press, New York, 1983, pp. 93--128.
MR 759907 (86f:35160).

\bibitem {k93}\textsc{C.E. Kenig, C. Ponce, L. Vega}, \textit{Well-posedeness
and scattering results for the generalized Korteweg-de Vries Equation via the
Contraction Principle,} Comm. Pure and Appl. Math. \textbf{46, }(1993), 527-620.

\bibitem {sulesulem}\textsc{C.Sulem, P.L.Sulem,}\textit{\ The Nonlinear
Schr\"{o}dinger Equation, }Springer New York (1999)

\bibitem {tsu70}\textsc{M.Tsutsumi,\ T.Mukasa,} Iino, Riichi, \textit{On the
generalized Korteweg--de Vries equation}, Proc. Japan Acad. Volume 46, Number
9 (1970), 921-925.\textit{\ }

\bibitem {We86}\textsc{M.I. Weinstein,} \emph{Lyapunov stability of ground
states of nonlinear dispersive evolution equations}, Comm. Pure Appl. Math.
\textbf{39} (1986), no.~1, 51--67.
\end{thebibliography}
\end{document}